\documentclass[finallayout,an,fleqn]{w-art}
\usepackage{times}
\usepackage{w-thm}
\theoremstyle{plain}

\theoremstyle{definition}

\usepackage[]{graphicx}
\chardef\bslash=`\\ 

\def \sgra  {Sgr~A*}
\def \Rsch  {R$_{Sch}$}
\def \etal  {et~al.~}
\def \masperyr {mas yr$^{-1}$}
\def \kms   {\ifmmode {{\rm km~s}^{-1}} \else {km~s$^{-1}$} \fi}
\def \Msun  {M$_\odot$}
\newbox\grsign \setbox\grsign=\hbox{$>$} \newdimen\grdimen \grdimen=\ht\grsign
\newbox\laxbox \newbox\gaxbox
\setbox\gaxbox=\hbox{\raise.5ex\hbox{$>$}\llap
     {\lower.5ex\hbox{$\sim$}}}\ht1=\grdimen\dp1=0pt
\setbox\laxbox=\hbox{\raise.5ex\hbox{$<$}\llap
     {\lower.5ex\hbox{$\sim$}}}\ht2=\grdimen\dp2=0pt
\def\gax{\mathrel{\copy\gaxbox}}
\def\lax{\mathrel{\copy\laxbox}}

\begin{document}
\DOIsuffix{theDOIsuffix}
\Volume{324}
\Issue{S1}
\Copyrightissue{S1}
\Month{01}
\Year{2003}
\pagespan{3}{}
\Receiveddate{3 March 2003}
\Reviseddate{}
\Accepteddate{}
\Dateposted{}



\title[Position, Motion \& Mass of \sgra]{The Position, Motion, and Mass of \sgra}


\author[Reid \etal]{Mark J. Reid \footnote{Corresponding
     author: e-mail: {\sf mreid@cfa.harvard.edu}, 
     Phone: +1\,617\,495\,7470,
     Fax: +1\,617\,495\,7345}\inst{1}} 
     \address[\inst{1}]{Harvard-Smithsonian Center for Astrophysics,
        60 Garden St., Cambridge, MA 02138, U.S.A.}
\author[]{Karl M. Menten\inst{2}}
     \address[\inst{2}]{Max-Planck-Insititut f\"ur Radioastronomie,
        Auf dem H\"ugel 69, D--53121 Bonn, Germany}
\author[]{Reinhard Genzel\inst{3}}
      \address[\inst{3}]{Max-Planck-Insititut f\"ur extraterrestriche Physik,
        Giessenbachstrasse, D--85748 Garching, Germany}
\author[]{Thomas Ott\inst{3}}
\author[]{Rainer Sch\"odel\inst{3}}
\author[]{Andreas Brunthaler\inst{2}}

\begin{abstract}
   We report progress on measuring the position of \sgra\ 
on infrared images, placing limits on the motion of the central
star cluster relative to \sgra, and measuring the proper motion 
of \sgra\ itself.  The position of \sgra\ has been determined
to within 10~mas on infrared images.  To this accuracy, the gravitational
source (sensed by stellar orbits) and the radiative source (\sgra)
are coincident.  Proper motions of four stars measured both in
the infrared and radio indicate that the central star cluster
moves with \sgra\ to within 70~\kms.  Finally, combining stellar 
orbital information with an upper limit of 8~\kms for the 
intrinsic proper motion of \sgra\ (perpendicular to the Galactic plane), 
we place a lower limit on the mass of \sgra\ of $4\times10^5$~\Msun.
\end{abstract}
\maketitle                   





\section{Introduction}

The precise position and proper motion of \sgra\ are of
fundamental importance in order to understand the nature of the 
super-massive black hole (SMBH) candidate and its environment.  
Unfortunately, \sgra\ lies behind about 30~mag of visual 
extinction, and currently it can only be detected in the radio, 
infrared, and x-ray bands.   While its radio emission is easily 
detected, the same cannot be said for its infrared and x-ray 
emission.  In both of these wavebands, emissions from nearby 
(in angle) stars make it difficult to isolate and measure the 
emission from \sgra.  Only with positions accurate 
to $\sim10$~milli-arcseconds (mas) can one confidently separate 
\sgra\ from confusing stellar sources and determine its 
spectral energy distribution and time variations.

Stellar proper motions, accelerations, and even orbits are 
now being determined to high accuracy at infrared wavelengths, 
and the position of the central {\it gravitational} source 
(presumably \sgra) can be measured to mas accuracy.  
If \sgra\ is indeed a SMBH, then the gravitational
source, inferred from stellar orbits, 
and the radiative source, directly seen in the radio band, 
should coincide to within $\sim 10$ Schwarzschild radii 
($10 R_{sch} \approx 0.08~{\rm mas} \approx 10^{13}$~cm for \sgra).  
Thus measuring the position of \sgra\ in the infrared to 
sub-mas levels is of fundamental importance in testing the 
SMBH paradigm.

The apparent proper motion 
of \sgra\ directly determines the sum of the angular
rotation speed of the Sun about the Galactic center,
$(\Theta_0+ V_\odot)/R_0$, and any peculiar motion of \sgra\ 
($V_{SgrA*}$) with respect to the dynamical center
of the Galaxy.  Thus, \sgra's proper motion can
provide a direct measurement of Galactic rotation.
In addition, the combination of stellar motions
and an upper limit on the motion of \sgra\ itself can
yield a strong lower limit to the mass of the SMBH candidate.

This paper reports recent progress on locating \sgra\ on
infrared images, measuring its proper motion, and placing a 
lower limit on its mass.

\section{Previous Results}

Menten \etal (1997) detected SiO maser emission from red giant
and supergiant stars within 12 arcsec of \sgra.  Since the
maser emissions originate from within about five stellar radii
of the host star, they can serve to precisely locate the star at
radio wavelengths relative to the strong radio source \sgra.
Also, the SiO maser stars are very bright at infrared wavelengths, and
one can use the radio positions of two or more stars to calibrate
the infrared plate scale and rotation and then align the infrared
image with the radio image containing \sgra.  Following this
method, Menten \etal located \sgra\ on a $2~\mu$m wavelength
image to an accuracy of 30~mas ($1\sigma$). 
No source of emission was seen at the position (see Fig.~1) 
of \sgra, and an upper limit of 9~mJy (de-reddened) was established.

\begin{vchfigure}
\includegraphics[width=0.9\textwidth,height=13.5cm]{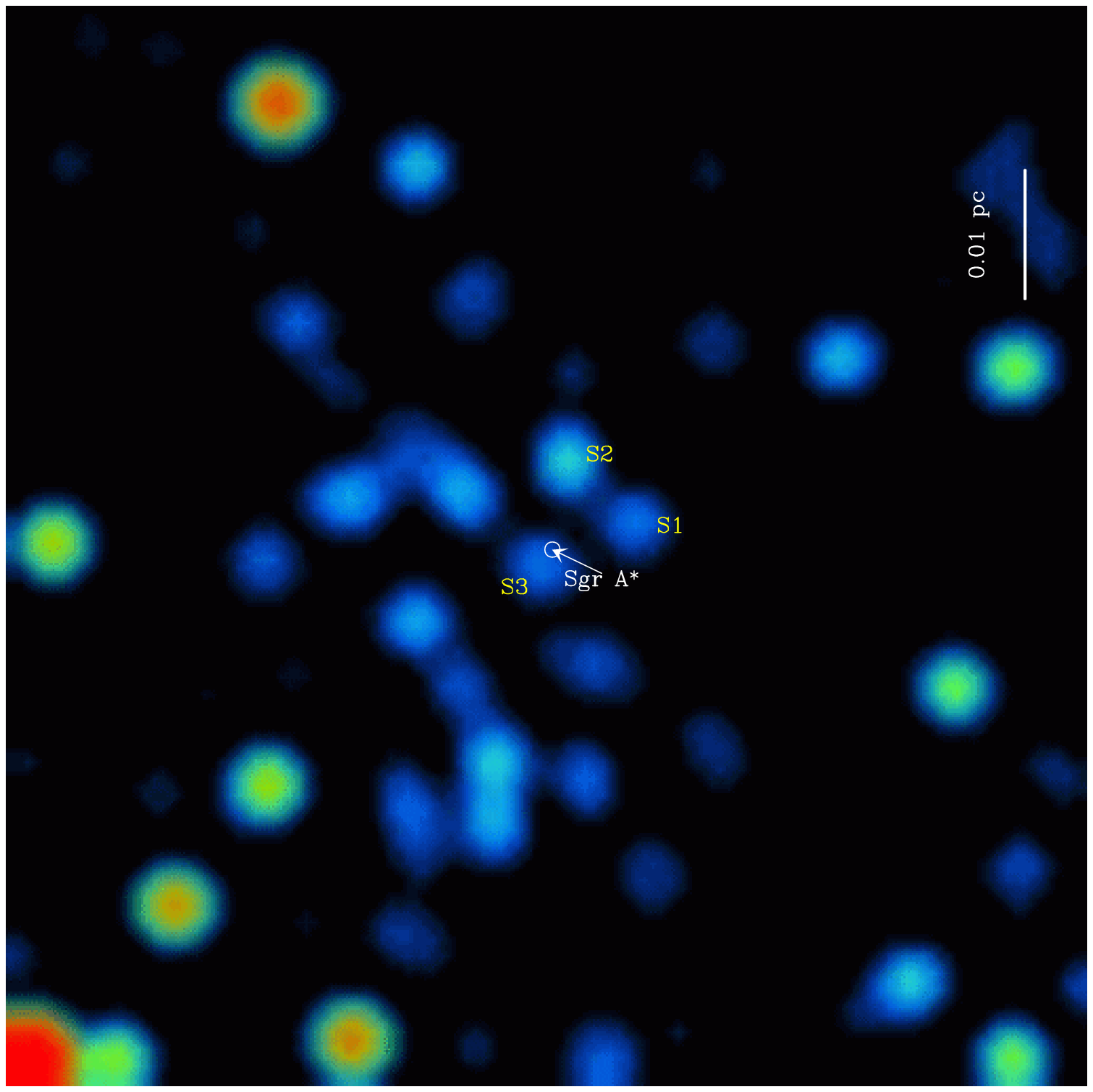}
\caption{Location of \sgra\ on a July 1995 $2\mu$m wavelength image of
the inner 2 arcsec of the Galactic center, adapted from 
Menten \etal (1997) by Reid \etal (2003).  The circle centered at the
position of \sgra\ has a radius of 15~mas, corresponding to a 
$1\sigma$ position uncertainty.}
\label{fig:1}
\end{vchfigure}

Reid \etal (1999) and Backer \& Sramek (1999) published observations
of the apparent proper motion of \sgra.  Both papers show that
\sgra\ appears to move toward the south-west along the Galactic plane
at about 6~\masperyr.  This is consistent with the angular rotation
rate of the Sun in its 220~Myr period about the Galaxy.   
Removing the effects of the Sun's orbit yields an upper limit
to the peculiar motion of \sgra\ of about 20~\kms.  Reid \etal
interpreted this upper limit to indicate that the mass of \sgra\
exceeds $\sim10^3$~\Msun, ruling out any stellar source.

\section{Recent Advances}

\subsection{The Infrared Reference System}

Recently, a wide-format infrared camera (CONICA; Lenzen \etal 1998) 
with an adaptive optics assisted imager (NAOS; Rousset \etal 2000) 
was installed on one of the ESO 8.2-m VLT telescopes.  This has 
produced excellent data for diffraction-limited imaging.  Very
deep images of the Galactic center with a field of view of 28~arcsec 
were taken with these instruments at $2~\mu$m wavelength early in 2002.  
These images proved to be of excellent quality.

Over the past seven years, radio frequency observations of SiO
maser sources in the Galactic center have been conducted with
the NRAO VLBA and VLA.  Both telescopes have been used to measure
positions and proper motions of SiO maser stars with accuracies of
about 1~mas and 1~\masperyr, respectively, relative to \sgra.
Seven maser stars within 15~arcsec of \sgra\ have now been measured 
to these accuracies (Reid \etal 2003).
By combining the radio positions of seven maser stars with their
apparent positions on the new VLT images, we could
determine the infrared plate scales and rotations with high accuracy.
After aligning the radio and corrected infrared images, we found
the residual differences in the maser star positions to be about 6~mas. 
This verified that the new infrared images have very small distortions 
across the field of view and allowed us to determine the position of
\sgra\ with a 10~mas ($1\sigma$) uncertainty (Reid \etal).

\begin{vchfigure}
\includegraphics[width=0.9\textwidth,height=13.5cm]{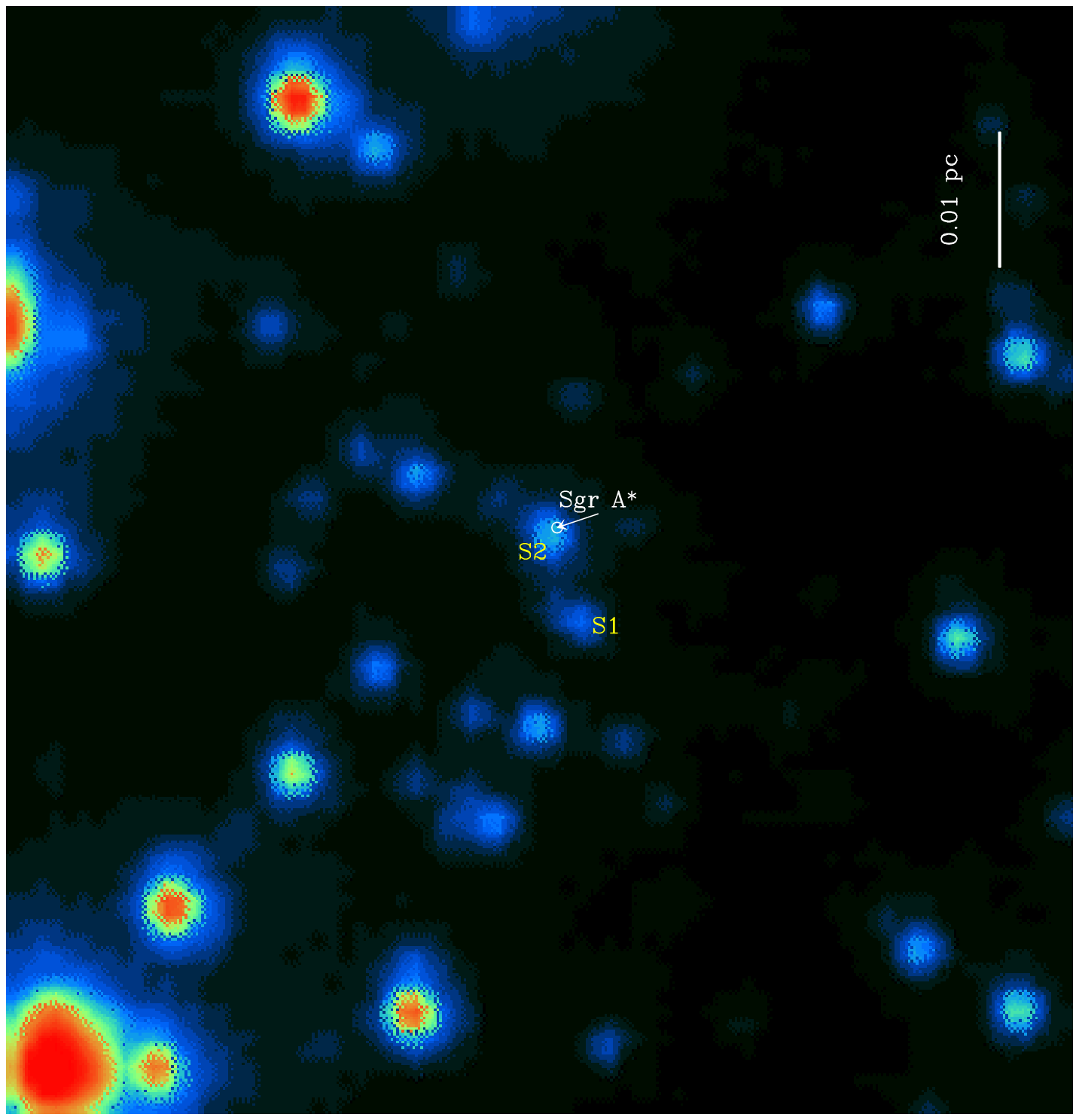}
\caption{Location of \sgra\ on 2 May 2002 on an infrared ($2~\mu$)
image of the inner 2 arcsec of the Galactic center, adapted from  
Reid \etal (2003). The circle centered at the
position of \sgra\ has a radius of 10~mas, corresponding to a 
$1\sigma$ position uncertainty.  The star S2 was near pericenter
in its orbit about \sgra\ when this image was taken
(Sch\"odel et al 2002; Ghez \etal 2003)
}
\label{fig:2}
\end{vchfigure}

Figure 2 shows a $2~\mu$m wavelength image taken on 2 May 2002, with
the position of \sgra\ and two nearby stars indicated.  
At this time, the fast moving star S2 was near pericenter in its orbit 
about \sgra\ (Sch\"odel \etal 2002).  The proximity of S2 to \sgra\ 
at this time (16~mas) precludes any significant measurement of the 
flux density of \sgra. 
The location for \sgra\ is within 10~mas of the gravitational source 
inferred from orbital solutions for the star S2 (Sch\"odel \etal; 
Ghez \etal 2003).  Thus, the radiative source (\sgra) 
and the gravitational source of the SMBH candidate are co-located to 
within about 1000~\Rsch.

Previous infrared proper motions have been
{\it relative} motions, with the motion reference defined
by setting the average of large numbers of stellar motions to zero.
We have compared the radio and infrared proper motions directly in order to
transfer the infrared motions to a reference frame tied to \sgra\ 
(Reid \etal 2003).
In principle, one can make this reference frame transfer using
a single star with well determined motions.
However, we chose to average the results from the four SiO maser stars 
within 10~arcsec of
\sgra\ that have measured proper motions both in the radio and infrared.
The unweighted mean difference (and standard error of the mean) of these
stars is $0.84\pm0.85$~\masperyr\ toward the east and 
$-0.25\pm0.96$~\masperyr\ toward the north.
Since 1~\masperyr\  corresponds approximately to 38~\kms (for a 
distance to the Galactic center of 8.0~kpc), we conclude that
the central star cluster moves with \sgra\ to within about 
40~\kms per coordinate axis, or within about 70~\kms for a
3-dimensional motion.

\subsection{Proper Motion of \sgra}

The {\it apparent} proper motion of \sgra, with respect
to extragalactic radio sources, was measured with the VLBA by 
Reid \etal (1999).  In that program, \sgra\ was used as a
phase reference to calibrate the interferometer phases for
two compact radio sources.  The location of these sources is
shown in Figure~3.   Relative to an extragalactic source,
\sgra\ would be expected to move toward the south-west,
mostly along the Galactic plane (as depicted in Fig.~3),
owing to the orbit of the Sun about the Galactic center.

\begin{figure}
\includegraphics[width=\textwidth]{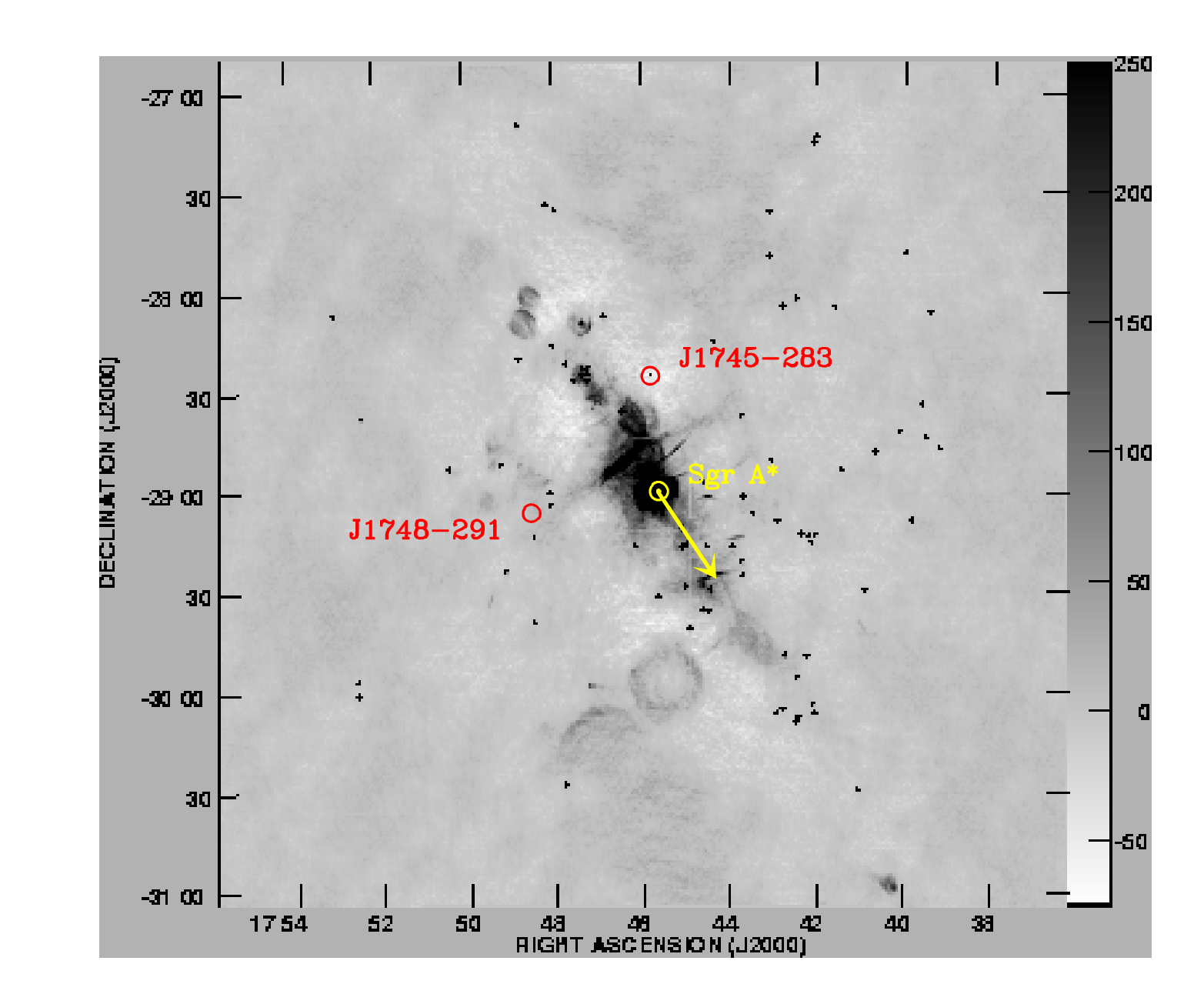}
\caption{Positions of \sgra\ and two compact extragalactic
radio sources superposed on a 90~cm wavelength image of the
Galactic center region made with the VLA by LaRosa \etal
(2000).  The expected motion of \sgra, owing to 
the orbit of the Sun about the Galactic center, is indicated
by the arrow.}
\label{fig:3}
\end{figure}

Figure 4 shows the position residuals of \sgra\ relative to
the compact extragalactic source J1745--283.  The positions
for 1995 through 1997 are from Reid \etal (1999).  Those
for 1998 through 2000 are new measurements.  As one can
see, the apparent motion of \sgra\ continues along the
Galactic plane.  
The dominant term in the apparent motion of
\sgra\ comes from the orbit of the Sun.  This can be decomposed
into a circular motion of the local standard of rest, 
$\Theta_0/R_0 \approx 220~\kms/8.0~{\rm kpc}$
(see Reid \etal 1999 for details), and the peculiar motion of the
Sun, $V_\odot/R_0\approx 20~\kms/8.0~{\rm kpc}$.  
Removing these terms from the observed
proper motion, yields estimates of the peculiar motion of
\sgra. 

\begin{vchfigure}
\includegraphics[width=0.66\textwidth,height=14cm]{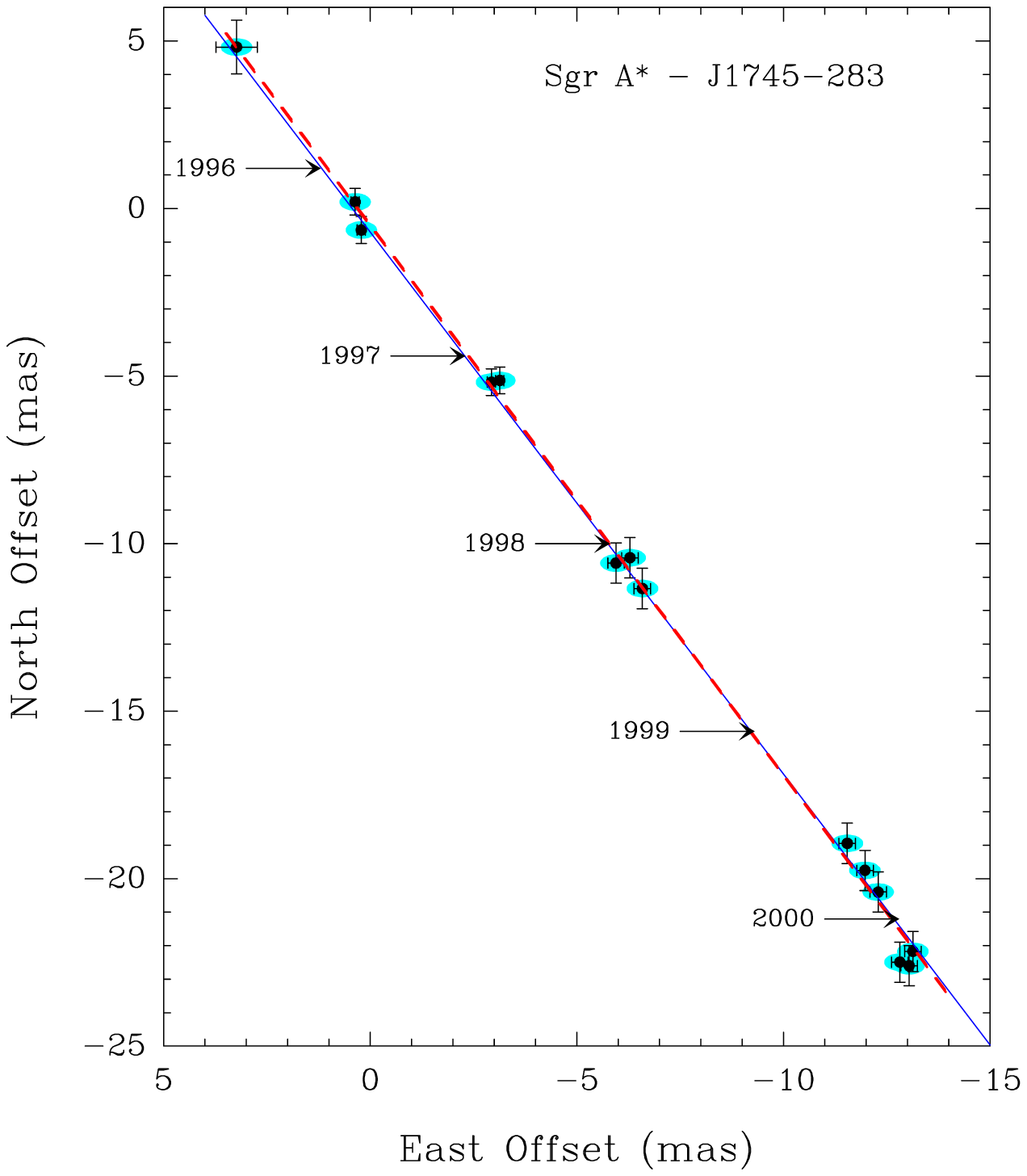}
\caption{Position residuals, with $1\sigma$ error bars, 
of \sgra\ relative to J1745--283 
on the plane of the sky. Each measurement is indicated with
an ellipse, approximating the apparent scatter broadened size
of \sgra\ at 43~GHz. The dashed line is the variance-weighted
best-fit proper motion, and the solid line gives the orientation
of the Galactic plane.  The expected position of \sgra\ at the
beginning of each calendar year is indicated.}
\label{fig:4}
\end{vchfigure}

While we currently do not know the component of $\Theta_0+V_\odot$ 
in the plane of the Galaxy to better than about 
10 to 20~\kms, we do know the component {\it perpendicular} to 
the Galactic plane to better than 1~\kms.
Since the circular motion of the LSR is, by definition,
entirely in the plane of the Galaxy, the only contribution
to the apparent motion of \sgra\ perpendicular to the plane of 
the Galaxy is the Z-component of the Sun's peculiar motions, ${Vz}_\odot$.
This component can be estimated by averaging the motions of 
very large numbers of stars in the Solar neighborhood, which should  
directly indicate $-Vz_\odot$.  An estimate, using
the Hipparcos database, indicates $Vz_\odot=7.16\pm0.38$~\kms\ toward
the north Galactic pole (Dehnen \& Binney 1998). 
After removing this contribution to the apparent motion of 
\sgra\ perpendicular to the plane of the Galaxy, we arrive at 
an estimate of $5\pm3$~\kms for this component of \sgra's peculiar motion. 
This result significantly improves on the limits given by
Reid \etal (1999) and Backer \& Sramek (1999).

\section{The Mass of \sgra}

From infrared observations of stellar orbits (Sch\"odel \etal 2002,
Ghez \etal 2003), we know that a mass of $\approx 3\times10^6$~\Msun\ 
is contained within a radius of $\approx100$~AU.  With this information,
and an upper limit on the Z-component of the velocity of \sgra, 
$V_z$, for which we adopt 8~\kms, 
one can estimate a lower limit to the mass of \sgra.   

The basic parameters of the problem are the total enclosed mass, 
$M_{enc}(R)$, including a possible SMBH and stars with typical individual 
mass, $m$, that are enclosed within a radius, $R$, and an upper limit 
on the Z-component of the velocity, $V_z$, of a ``test'' object 
(\sgra\ in our case) of mass $M$. 
In the past, two limiting cases of mass estimators have been discussed
for this problem: equipartition of kinetic energy (see Backer \& Sramek 1999) 
and momentum (see Reid \etal 1999).  
Equipartition of kinetic energy implies that
$$MV^2 \sim mv^2~~,\eqno(1)$$ 
where $v^2\approx GM_{enc}(R)/R$ is a characteristic stellar 
velocity at radius $R$ (which must be great enough so that the mass
in stars exceeds that of the the test object, i.e., $M_{enc}(R)\gax2M$). 
Equipartition of energy is both theoretically and observationally well 
founded for the case of stellar clusters. 
However, for the case of a dominant central mass, which 
could greatly exceed the total mass of stars (within a given radius), 
Reid \etal  argued that one would be dealing with true orbits and that 
equipartition of momentum would then be appropriate:
$$MV \sim mv~~.\eqno(2)$$
It turns out that both estimators are correct, but for answering 
different questions.  

If one asks what is the expected velocity of  
a SMBH that is perturbed by close passages of stars which 
orbit it, then the momentum equation~(2) applies.  This is almost
surely the case for \sgra\ and nearby stars such as S1 and S2.   
For star S2, which has a mass $m$ of
$\approx15$~\Msun\  (Ghez \etal 2003), during pericenter passage
$v\approx 6500$~\kms\ and Eq.~(2) implies that 
one would expect a $3\times10^6$~\Msun\  SMBH's peculiar motion to be  
$V \sim 0.03~\kms~~.$
Following this approach one can calculate an extremely conservative
lower limit for \sgra's mass.  While this is valid, it is not an optimum
estimate. 

Alternatively, if we ask for the minimum mass of a central object
which does {\it not} totally dominate the enclosed mass, $M_{enc}(R)$, within
a given radius $R$, and which complies with the observed velocity
limit, then we have a different case.  For this case, where the
enclosed mass in stars within $R$ is comparable to or exceeds
the mass of \sgra, $M$, equipartition of kinetic energy should apply.
When evaluating Eq.~(1) one must use velocities for stars with
radii near $R$.
Conceptually, as the velocity limit for \sgra\ improves,
the estimated mass limit increases quadratically in $V$.  
This continues until the estimated
mass dominates over the stellar component and our assumption
is violated.  At this point, however, one has already ascribed most of
the enclosed gravitational mass to \sgra.

A recent paper by Chatterjee, Hernquist \& Loeb (2002) analyzes
our mass estimation problem in a manner similar to that described above.
They assume a black hole at the
center of a stellar cluster, which is distributed in space according
to a Plummer profile with a characteristic scale $a$.
The {\it minimum} black hole mass occurs for $a$ approximately 
equal to the radius, $R$, within which the enclosed mass is measured.  
In this case, the mass estimator (their equation 42) can be simplified to 
the following:
$$M \sim {G M_{enc}(R) m \over V_z^2 R}~~,\eqno(3)$$
provided $V_z^2>Gm/R$, which is met   
for $V_z=8$~\kms, $m=1$~\Msun, and $R=100$~AU.  
Then for the observed $M_{enc}(R)=3\times10^6$~\Msun, 
Eq.~3 gives a lower limit to the mass of \sgra of 
$M \gax 4\times10^5$~\Msun.

Our lower limit to the mass of \sgra\ is now within about
a factor of 10 of the total mass required by recent
IR data.  Since the uncertainty in the proper motion ($\sigma_V$)
can decrease with the spanned observing time ($T$) as 
$\sigma_V\propto T^{-3/2}$, and the lower limit to the mass 
of \sgra\ scales as $\sigma_V^{-2}$ (until the limit approaches
the total enclosed mass), we can expect improvement in the limit
for $M \propto T^3$ over the next few years.   
When we reach a motion limit of 1 to 2~\kms for \sgra, then 
essentially all of the mass sensed
gravitationally by stellar orbits must come from \sgra\ itself.
Should future VLBI measurements at $\lax1$ mm wavelength show that 
the intrinsic size of \sgra\ is $\lax0.1$~AU, then we may be in a
position to conclude
that for \sgra\  most of the mass required for a SMBH is contained within
a few $R_{Sch}$!

%

\end{document}